# Contextual AI Journaling: Integrating LLM and Time Series Behavioral Sensing Technology to Promote Self-Reflection and Well-being using the MindScape App

Subigya Nepal
Dartmouth College
Hanover, New Hampshire, USA

Arvind Pillai
Dartmouth College
Hanover, New Hampshire, USA

William Campbell
Colby College
Waterville, Maine, USA

Talie Massachi
Brown University
Providence, Rhode Island, USA

Eunsol Soul Choi
Cornell Tech
New York, New York, USA

Orson Xu
Massachusetts Institute of Technology
Cambridge, Massachusetts, USA

Joanna Kuc
University College London
London, UK

Jeremy Huckins
Biocogniv Inc
Burlington, Vermont, USA

Jason Holden
University of California, San Diego
La Jolla, California, USA

Colin Depp
University of California, San Diego
La Jolla, California, USA

Nicholas Jacobson
Dartmouth College
Hanover, New Hampshire, USA

Mary Czerwinski
Microsoft Research
Redmond, Washington, USA

Eric Granholm
University of California, San Diego
La Jolla, California, USA

Andrew T. Campbell
Dartmouth College
Hanover, New Hampshire, USA

## ABSTRACT
MindScape aims to study the benefits of integrating time series behavioral patterns (e.g., conversational engagement, sleep, location) with Large Language Models (LLMs) to create a new form of contextual AI journaling, promoting self-reflection and well-being. We argue that integrating behavioral sensing in LLMs will likely lead to a new frontier in AI. In this Late-Breaking Work paper, we discuss the MindScape contextual journal App design that uses LLMs and behavioral sensing to generate contextual and personalized journaling prompts crafted to encourage self-reflection and emotional development. We also discuss the MindScape study of college students based on a preliminary user study and our upcoming study to assess the effectiveness of contextual AI journaling in promoting better well-being on college campuses. MindScape represents a new application class that embeds behavioral intelligence in AI.

## CCS CONCEPTS
• **Human-centered computing** → **Ubiquitous and mobile computing**; • **Applied computing** → **Health informatics**.

## KEYWORDS
Passive Sensing, Large Language Models, Journaling, Self-reflection, Behavioral Sensing, Mental Health, Well-being, AI, Smartphones

## 1 INTRODUCTION
The significance of struggles with mental health among college students is becoming increasingly apparent, impacting students' academic performance, social engagement and overall personal development. Research, including findings from the American College Health Association (ACHA)–National College Health Assessment, highlights a concerning prevalence of anxiety, depression, and related issues among students [4, 7, 33, 35]. Students face a range of pressures, from academic challenges to social and personal hurdles, which affect not only their mental health but also their emotional resilience and personal growth [22, 44–46]. While traditional mental health interventions administered by clinicians do provide personalized and context-specific support, emerging technologies present an opportunity to extend this support, making it more readily available, automated, and able to potentially overcome considerable institutional barriers. In addition, there is a need for innovative solutions that align with the digital habits of today's students. We propose a novel study, *MindScape*, that integrates the traditional practice of journal writing with mobile technology and large language models (LLM) [34] to create a contextually-aware journaling application. The MindScape Android application benefits from on-device sensors and data to provide insights into the user's daily life. It tracks aspects such as physical activity, social interactions, and location to form an understanding of the context in which the individual operates. By analyzing these data in real-time, the app can provide personalized, context-sensitive journaling prompts designed to provoke thought and reflection. The prompts aim to remind users to introspect and commit time to digitally record their thoughts, thus establishing regular self-reflection habits that are contextualized by their daily lives. MindScape represents a novel application class that incorporates behavioral intelligence into AI. We believe that integrating time-series data obtained from mobile phones and wearables, capturing real-time behaviors and patterns of users, with the capabilities of Large Language Models (LLMs) will give rise to a new category of AI applications driven by mobile sensing.

Journaling has long been recognized as a potent tool for self-reflection, enabling individuals to externalize thoughts, consolidate disjointed experiences, and identify patterns in their behavior and emotional states. This practice of regular introspection has been linked to a range of psychological benefits, from reducing distress symptoms to enhancing overall wellbeing [17, 43]. In this study, we explore the potential gains realized through the inclusion of personalization and context-awareness in journaling. The inclusion of personalization and context-awareness in journaling is not merely an innovation for convenience's sake. Instead, it addresses certain inherent limitations in human introspection and memory recall abilities. People may not readily identify certain behavioral patterns or come to particular conclusions about their daily lives without some form of guidance or external input. This is where personalized and context-aware prompts can be valuable, as they may highlight aspects of users' lives they may have overlooked. Additionally, human memory recall can be biased towards more recent experiences, sometimes at the expense of equally significant past events. Context-aware journaling can help counteract this limitation by bringing forward relevant circumstances, events, or feelings from different timeframes in the users' life. Lastly, by addressing these user limitations, personalized and context-aware journaling could not just improve the process of journaling, but also potentially enhance the mental health benefits associated with this practice.

Herein lies the novelty of our approach: using mobile sensing to capture behavioral data that reflects the user's context and emotional state, and employing an LLM to generate journaling prompts that are highly relevant to the user's current contextual situation and surroundings. In addition to the context-aware journaling prompts, our study app also triggers context-aware check-ins (such as, *"Your morning seemed to include more than just tapping screens – a bit of chitchat too!"*) that are delivered four times daily to encourage quick moments of reflection as opposed to longer-form journaling. The daily check-ins are short and simple texts relevant to what the user is experiencing at the time they are issued. Users can answer these contextual check-ins with a simple thumbs up or thumbs down button – they are designed for fast, low burden response. This relevance of check-ins can increase the user's engagement and attachment with the journaling app, and similarly, the context-aware nature of the journaling prompt can make their entries more meaningful, potentially amplifying the mental health benefits they receive from journaling. Furthermore, the MindScape journaling app integrates additional contextual factors such as students' mood while journaling, their academic stress levels, and temporal variables like weekdays or weekends. Early in our development, we conducted a qualitative user study with undergraduate students to understand their journaling habits and preferences. Insights from this study, revealing students' desires for personalized, context-aware prompts aimed at fostering reflection on daily experiences, significantly influenced the app's design. We believe our holistic approach allows for a more tailored and responsive tool, capable of providing meaningful support in the unique, often high-pressure, fast paced environment of college life [31]. Some of our research questions are as follows:

- **RQ1:** *How does integrating mobile sensing and AI-driven journaling affect college students' mental well-being and personal growth?*
- **RQ2:** *How do context-aware journaling prompts, informed by mobile sensing data, contribute to enhancing the depth, detail, and insightfulness (key measures of quality) of students' reflective journaling practices?*
- **RQ3:** *Is the integration of mobile sensing and AI-driven journaling a feasible and acceptable approach for college students, and how do they perceive its impact on self-awareness and emotional regulation?*
- **RQ4:** *How do college students perceive the overall functionality and experience of the MindScape app?* This includes evaluating what aspects are most effective and identifying areas for improvement in the context of a technology-driven, personalized journaling experience.

The MindScape study is designed to enhance the classic benefits of journaling by utilizing the latest advancements in LLMs to provide an unobtrusive, effective tool for users to manage their wellbeing and growth. This approach is closely aligned with the Human-Computer Interaction (HCI) community's interests, highlighting the significance of AI in enriching user-centric digital experiences. Bridging into Ubiquitous Computing (UbiComp), our research focuses on integrating these technologies into everyday routines. Our goal is for this tool to offer benefits and support and assist students in developing lasting self-reflection and emotional mindfulness skills. We hope that this study will contribute significantly to the ongoing dialogue in HCI and UbiComp, particularly regarding the seamless integration of technology to enhance personal well-being, offering a comprehensive view of its practical application and user impact.

## 2 RELATED WORK

Journaling is a reflective practice where individuals record their thoughts, feelings, and experiences. The act of journaling promotes self-awareness [1, 47], processing of emotions [42], and cognitive organization of experiences [43]. Studies have consistently shown that journaling can improve mood, provide stress relief, and overall, enhance mental well-being [24, 32, 43]. As mobile devices and computers become more prevalent, they have reshaped the practice of journaling. The transition to digital journaling platforms brings conveniences that traditional paper-based methods lack. These include enhanced accessibility, ensuring that users can journal anytime and anywhere, heightened privacy—as entries are secured behind digital safeguards—and the ability to enrich journal entries with multimedia elements.

Journaling can be prompted or unprompted. Unprompted journaling allows for free expression without specific guidelines, giving users freedom to explore their thoughts and feelings. In contrast, prompted journaling uses specific questions or suggestions to guide the journaling process, providing a structure that can help focus and inspire the user. Such prompts are designed to encourage self-reflection, personal growth, and exploration of various topics and experiences. Several digital journaling platforms offer a wide range of prompts to initiate the writing and reflection journey, providing daily reminders to ensure users stay on track with their journaling. This approach can be particularly helpful for users who are new to journaling or those looking to explore new areas of self-discovery and creativity. However, most prompted journaling applications rely on generic prompts not tailored to the user's situation. Several studies demonstrate that question prompts are one of the main factors positively affecting reflection quality [12, 13, 19, 20]. Thus, generic prompts, while useful, may reduce reflection quality due to their broad nature [3, 39].

Our study focuses on context-aware journaling, where journaling prompts are derived from behavioral data collected via smartphones. This approach enhances traditional journaling by offering prompts that closely align with users' daily experiences and mental states. By using mobile sensing technology, capable of tracking activities, sociability, locations, and app usage, we generate dynamic prompts that reflect the nuanced aspects of an individual's life. This approach differs from previous studies that have explored a broader range of personal informatics systems for reflection [9, 15], by integrating these insights into the journaling process. For example, Kocielnik et al. [28] leverage mobile based step count for reflection on activity level whereas Bakker and Rickard [6] use the MoodPrism app to help in mood tracking. Our method aims to mirror the reflective goals of various fitness and sleep apps and to offer deeper insights into users' lifestyles and emotional patterns through personalized journaling. In addition, our study uses a wide range of contextual cues to facilitate journaling, a feature that sets it apart even from its closest counterparts like Apple's journal application [2]. While Apple's offering leverages contextual data such as photos and location to generate prompts, it is positioned within the iOS system and does not fully tap into the exhaustive range of cues that our study incorporates on Android devices. For example, our approach extends beyond conventional context-awareness to include an amplified set of signals like digital habits, which includes screen time, social, entertainment, and communication app usage. Our methodology also captures levels of social interaction, including in-person conversations, calls, and text message exchanges.

Further, our study integrates both sleep information, such as duration and timing, as well as physical fitness metrics like activity levels, distance travelled, and time spent at the gym. Our study also considers location-based semantics like time spent in a cafeteria, Greek spaces, and other similar locations. This comprehensive approach sets our study apart by providing a more nuanced and detailed context for generating personalized journaling prompts. Our study also leverages LLM capabilities to enable the creation of intelligent, personalized journaling prompt. AI-driven tools have been used in therapy chatbots, virtual agents, and behavior change systems, offering personalized advice and support [10, 14, 23, 25, 29, 30, 36, 41, 48–50]. These applications demonstrate the capacity of AI to understand and respond to a wide range of emotional and psychological states. Existing studies have leveraged LLMs for AI-mediated journaling [18, 26, 27]. However, to our knowledge, none of the existing studies have integrated objective and passively observed behavioral data into AI-mediated journaling. By using an LLM framework to analyze behavioral data and generate relevant journaling prompts, we aim to investigate the potential for a nuanced, data-driven augmentation of the journaling process. Our study seeks to reinforce the benefits of journaling, while simultaneously exploring the effectiveness of context-aware prompts for highly reflective self-expression. Through this unique approach, we aim to optimize the impact of personalized digital journaling.

## 3 METHODOLOGY

### 3.1 Study Design

We plan to enroll approximately 40 participants. This number is tailored to balance a sufficient sample size while accounting for technological constraints, particularly as the MindScape passive sensing app is only compatible with Android devices. We plan to target participants who already practice journaling, aiming to assess the added value of MindScape's contextual prompts for those experienced in reflective writing. We anticipate that journaling individuals may more readily engage with the app's features. Participants will be compensated, with a potential compensation of up to USD 130. The study includes initial onboarding surveys, weekly surveys (also known as Ecological Momentary Assessments (EMA)), and concluding surveys over 8 weeks. Our study excludes participants with high depression, identified by elevated PHQ-8 survey scores, to ensure safety, since the prompts are not manually screened for sensitive content. We additionally interviewed five students at Dartmouth College to gain insights into their preferences for a journaling application, grounding our approach in user-centered design principles. The study is approved by Dartmouth's Internal Review Board.

### 3.2 Mobile Sensing based Behavioral Data

The MindScape app automatically infers user activities, like movement and rest, analyzes conversation lengths, and gathers data on screen usage and location (see Table 1). This provides an integrated view of a user's daily patterns, social interactions, and digital habits. For example, the sensing data might reveal patterns in how often participants attend social functions, dine at campus facilities, or go to the gym. This information allows us to tailor the journaling prompts to align with the participant's current experiences and to support their emotional well-being. As part of gathering this data, we created a semantic map of the college campus, with locations such as dining areas and gyms marked, allowing the app to accurately infer the context of participants' activities. This allows for prompts to be customized, encouraging reflection on particular events of the day. The integration of the GPT-4 LLM enables the translation of this rich, multi-faceted behavioral data into personalized and contextually relevant journaling prompts and frequent check-ins that enhance positive introspection and participant engagement. All data collected are temporarily stored on the participant's phone and then securely uploaded to the MindScape cloud. We then leverage the GPT-4 model through OpenAI's API [38], allowing us to process the collected behavioral data and additional contexts to generate tailored prompts. Addressing potential concerns relating to participant privacy, we ensure all data sent for processing via OpenAI's GPT-4 model are de-identified and consist only of high-level metadata. This approach includes stripping any potentially personally identifiable information before the data is utilized to generate tailored prompts. We acknowledge that a locally hosted open-source model could offer an alternative to mitigate privacy concerns further, albeit with possible performance tradeoffs. In this study, our focus has been oriented towards understanding the potential and efficacy of this novel application of AI in journaling practices. Given this emphasis, we decided to utilize OpenAI's GPT-4 model for its robust performance and scalability capabilities.

Table 1: Behavioral Data Categories: Users are required to prioritize among the four behavioral data categories, each encompassing specific feature sets.

| Category | Signals/Features | Example Journaling Prompt Generated |
| --- | --- | --- |
| Physical Fitness | Physical activity (walking, running, and sedentary duration) | *Your running routine has really taken off! How's that influencing your day?* |
| | Distance travelled | |
| | Time spent at the Gym | |
| Sleep | Sleep duration | *Your sleep pattern has shifted recently. Could this change be affecting your daytime energy and focus?* |
| | Sleep schedule (start time and end time) | |
| Digital Habits | Screen time | *You've been clocking less screen time lately. What have you been doing instead that you've found rewarding or enjoyable?* |
| | App use (Freq. of social media, communication, & entertainment apps use) | |
| Social Interaction | Phone logs (incoming calls, outgoing calls, incoming SMS, outgoing SMS) | *Your call patterns are up; any conversations lately that brought a smile to your face?* |
| | In-person conversations (number and duration of conversations) | |
| | Number of significant places visited | |
| | Time spent at frats/sororities partying | |
| | Misc. locations (Time spent at leisure, social, study places, cafeteria & home) | |

## 3.3 Personalized Journaling Prompts

Upon installing MindScape app, participants will allow the app permission for data collection. Then, they will rank their journaling interests in four key areas — Social Interaction, Sleep, Digital Habits, and Physical Fitness. We identified these four key areas through interviews with students on campus (See Section 4). Because we collect many different types of data, we want to ensure the journaling prompts we provide are actually helpful to participants. Thus, we use these categories to identify what matters most to each individual participant. We also include the user's preferences (i.e., category ranking) in the prompt for GPT-4 [37] to generate more relevant journaling prompts. During their enrollment, each user enters their usual bedtime for both weekdays and weekends. Journaling notifications are triggered two hours before their reported bedtime. When a notification is tapped, participants are redirected to the app's journaling screen. There, they are first asked how their day was, followed with a one-minute breathing exercise, and finally, they are asked to write or record (i.e., audio) their journal entry. Only at this point can the participants see the personalized journaling prompt. Participants can also open the app and journal whenever they prefer. Note, the one-minute deep breathing exercise before journaling is based on findings that short relaxation techniques can improve mental clarity and emotional readiness [8, 51]. This step aims to help users transition to a reflective mood, enhancing their focus for more insightful journaling. It's intended to make the journaling process a calming, enriching routine. Figure 1 shows different screens of the application.

**Contexts** The GPT-4 prompt composition process incorporates several layers of contextual data:

- **Personal Priorities:** The user's preferences across the four journaling categories ensure that the journaling prompts mirror individual interests.
- **Prompt Variability:** The system ensures that new prompts are different from the previous two, generating diverse and engaging content.
- **Temporal Data Analysis:** Behavioral data from weekdays are contrasted with a 30-day historical average to establish context. On Saturdays, the app encourages users to reflect on general themes from the preceding week, rather than daily behaviors (for example, *"Recall a recent academic success. How did you achieve it and what did it teach you about your resilience or strategy?"*). Sundays are used for a comprehensive review including additional data points - such as Greek house attendance and sleep quality - to capture weekend patterns pertinent to college life. *Note: In the U.S., 'Greek houses' are fraternity or sorority residences, where social and organizational activities are hosted.*
- **Academic Calendar Awareness:** As the academic term structure influences stress, the current week of the term is considered during prompt generation, intending to offer supportive content during high-stress phases.

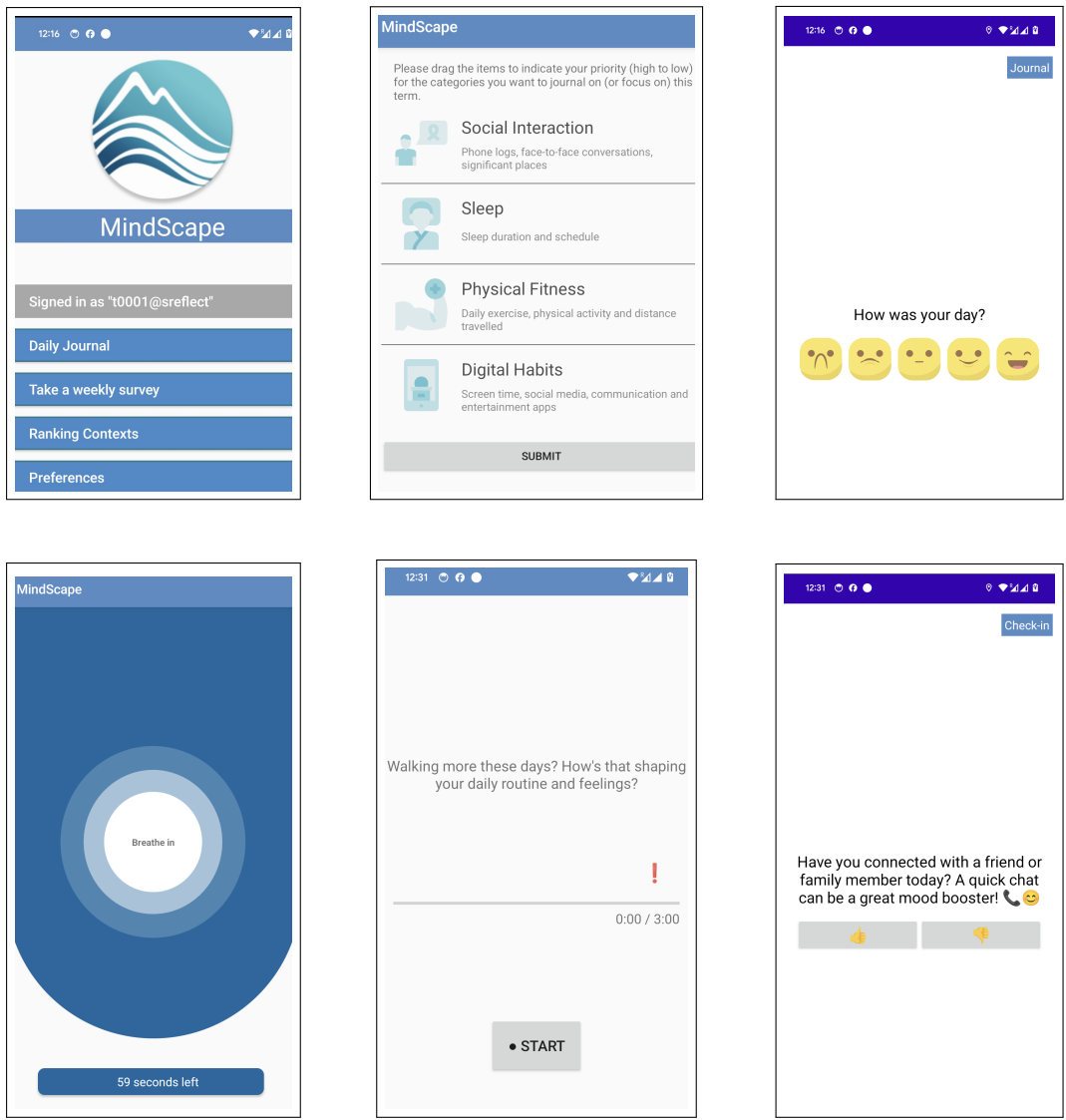

Figure 1: MindScape App Workflow: Users sign in, set preferences in four categories, and start journaling with a notification tap. Journaling workflow includes an emoji based mood assessment, a one-minute breathing exercise, followed by contextual prompts. The final screen showcases the daily context-aware check-in.

- **Mood Consideration:** If a participant reports a low mood, GPT-4 is prompted to offer journaling prompts that evoke self-compassion or gratitude—strategically fostering a nurturing journaling environment. By guiding users towards reflecting on aspects they are grateful for or encouraging kindness towards themselves, the hope is that these prompts can shift focus from negative thoughts to more positive, affirming ones. It is a strategic, evidence-based approach aimed at offering immediate emotional relief while contributing to long-term emotional well-being, resilience, and mental health [16].

Our methodology emphasizes customization, employing both user preferences and behavioral signals to empower participants in their reflective journaling practice. In Figure 2, we show how all these come together to form the input to the GPT-4 LLM.

### 3.4 Context-aware Check-ins

The check-ins are "micro context-aware nudges" based on users' data, and are answered with a quick thumbs up or thumbs down response. For example, *"Caught up with some calls and social apps this morning - digital world kept you busy, I bet!"*. The MindScape app offers such check-ins four times a day at 12.30 PM, 3.30 PM, 6.30 PM and 11 PM. These times are strategically selected to suit the daily rhythms of college students, ensuring the interaction remains brief and unobtrusive.

## LLM Input Prompt

### 1. System Prompt

You are an assistant that is helping create responses for a self-journaling application. Given the user's mood score (most important), sensing data, and previous responses, create a succinct self-journaling prompt. The intended user are undergraduate students at a top institution. The prompt must pinpoint a specific behavior or pattern that might need improvement or adjustment. Design the prompt in a way that forces users to do some type of interpretation and encourage them to respond to in their own words. For context use the user trends, mood score and the community stress index. Take note of how user is feeling (i.e., their mood) before crafting the prompt that would be appropriate for them to see. The response should be focused on the user priorities ranked from highest to lowest. The response should have a greater sense of personality, relationality, to solicit rich narratives. To mitigate against the "cold" feel of the response, create a "voice" in the response that would have some warmth, such as including thank you messages.

### 2. User Context

**[User Mood]**
User Mood Score (1 lowest, 5 highest): *<Self-reported mood>*
**[Previous Prompts]**
Previous Responses: *<List of previous journaling prompts>*
**[Community Stress Index]**
Community Stress Index (1 lowest, 5 highest)*:*
  *<Stress index generated based on current week in the term>*
**[User Priorities]**
Priorities [highest to lowest]*: <Self-reported priority>*

**[User Data]**
Today's Date: *<date>*
*<List of user trends>*
  Duration of incoming phone calls: increasing by 10%
  Walking duration: stable
  ….
  Screen time: decreasing by 12%
  Frequency of use of social apps: increasing by 20%

### 3. Prompt Optimization

**[Response Rules]**
1. Response should not use over-the-top words. Only conversational language.
2. Avoid repetitive response by using the context of the "Previous Prompts". Do not mention the same idea conveyed in Previous Prompts.
3. Do not provide a generic or offensive, argumentative, or mentally damaging response, instead be friendly and upbeat.
4. Do not mention specific data points (for example do not say "Your mood score was 1/5 for the week").
5. A suitable example would be "Your step count indicates that you have been more active today than the rest of the week. What activities did you engage in, and how did they make you feel?"
6. Do not write leading questions. The response should not be too open-ended. Make it more direct so that it can help direct thoughts better.
7. Have a conversational and qualitative response, not a clinical and quantitative one (for example do not say "I saw your sleep is deteriorating").
8. Should be less than 250 characters.
9. MUST respond with only the prompt, do not give any prefix such as "prompt" and do not use double quotes at the start and end of the response.
10. Avoid any repetitions that might be stigmatizing or feel exclusionary (such as, "if you have a partner"). Response should speak to both group and individual situations.
11. Since this is being used by young students, make the response more relatable, trendy, gen-z friendly.
12. Do not include any "tip: ", "question: " etc and do not use hashtags.

### 4. Strategy

If user reports low mood score (<3), the response should either elicit gratitude towards others or encourage self-compassion. Choose one randomly. In case of self-compassion, it should promote self-positivity and encourage users to reflect on positive attributes in their life rather than trying to solicit rich narratives. In case of gratitude, it should ask users to respond about things they are grateful for today. Ideally, it teaches users to shift their attention away from negative aspects of their lives and instead to learn to appreciate what they have and what others have done for them. This mindset shift counterbalances the "negativity bias" or one's tendency to pay more attention to negative stimuli in the environment.

## LLM Output

**[Gratitude]**
With academic term stress peaking, have you found moments to appreciate successes or kindnesses from your day?

**[Self-compassionate]**
Amid your busy days, you've been getting more sleep but also more screen time. How do you feel about this trade-off, and could there be a balance you'd like to aim for?

**[Standard]**
Given that you're feeling on top of the world today, how has interacting with your friends via texts and calls contributed to your vibe?

**Figure 2: Prompt Template for Weekday Journaling:** The input prompt to GPT-4 is composed of four parts: 1) System prompt 2) User context 3) Rules to optimize the prompt and 4) The strategy to generate the journaling prompt.

Each check-in is designed to incorporate the behavioral data gathered during the time period extending from the previous check-in up to the current one. For instance, the 3:30 PM check-in uses data collected from 12:00 PM to 3:30 PM, while the 6:30 PM check-in uses data gathered from 3:30 PM to 6:30 PM. This approach ensures that each check-in is responsive to the most recent behavioral data captured for the participant. The goal of these check-ins is to both increase the visibility of the app (as opposed to users seeing it just once a day for journaling) as well as to increase reflection on behavior through a casual, quick touchpoints.

## 4 RESULTS

### 4.1 Phase 1: User Study

We conducted qualitative user studies comprising in-depth interviews with five undergraduate students at Dartmouth College, with the aim of understanding their journaling habits, preferences, and reactions to personalized prompts potentially generated by the MindScape app. The students were aged between 18-24, consisting of three males and two females. These participants were sought through invitations extended by the team, targeting individuals who could provide important insights into the efficacy and impact of personalized journaling prompts within the context of university life. From these interviews, we obtained several key insights:

- **Journaling Preferences:** Students indicated a preference for using journaling as a means to tackle personal challenges or to meet specific objectives, such as improving gym routines. They favored insightful prompts and identified four main areas of interest: Social Interaction, Sleep, Physical Fitness, and Digital Habits.
- **Self-reports and Sensing-based Prompts:** Students were willing to provide self-reports at different times of day, suggesting that multiple daily check-ins and end-of-day journaling are feasible. They responded positively to location-specific prompts, such as those related to meals in the cafeteria or academic work in the library.
- **Mood based prompts:** The interviews highlighted a preference for contextually timed journaling prompts. For instance, during stressful periods like exams, students preferred prompts that focus on positive aspects and are easy to engage with.
- **Academic Stress:** The study confirmed our understanding of the academic stress patterns among students.

These insights have shaped the development of MindScape's prompting mechanisms to support users in engaging with their reflective practices in a meaningful way. For instance, we integrated additional contexts into the app such as academic calendar awareness, mood consideration, weekday vs weekend prompts, and personal priorities.

## 4.2 Phase 2: Expected Results from Larger-Scale Study

We are currently recruiting participants for the larger-scale study described above. To evaluate the impact of the MindScape app on student well-being and personal growth, we will use a combination of standardized surveys, participant feedback, and app insights. We anticipate positive changes in mindfulness, emotion regulation, mental well-being, personal growth, resilience, life satisfaction, social connections, self-reflection, and personal insight. For each of these areas, specific surveys have been chosen to measure changes effectively. For instance, the Five Facet Mindfulness Questionnaire [5] will assess changes in mindfulness, while the Emotion Regulation Questionnaire [21] will gauge emotion regulation. Similarly, surveys like Ryff's Scales of Psychological Well-being [40] and the Flourishing Scale [11] will help evaluate overall mental wellness. In this proof-of-concept phase, any level of change observed, or even its absence, will offer insights into user interactions with our app and its usability, crucial for HCI considerations. This initial study, focusing on students from a single campus with previous journaling experience and no mental health disorders, is as much about understanding user engagement as it is about quantifying outcomes. We are mindful that changes observed in this specific demographic may be less pronounced than in other groups, such as those with mental health disorders.

Our goal, therefore, is also to evaluate the effectiveness and user reception of the MindScape app, leveraging feedback on its functionality and the impact of its unique contextual prompts for guiding future enhancements. Our exit surveys will probe into aspects like app performance and users' experiences with contextual versus traditional generic prompts. To address our research objectives, we plan to undertake a multi-faceted analysis:

(1) By conducting a comparative review of wellbeing and personal growth survey responses collected before and at the end of the study, we aim to identify any positive shifts.
(2) We will solicit direct feedback on the app through questions regarding users' observations of changes, overall satisfaction, standard usability metrics, among other things.
(3) Additionally, we intend to apply Natural Language Processing (NLP) techniques to analyze the content of journal entries, offering insights into user engagement and the reflective outcomes of journaling.

This research lays the foundation for developing an automated system for creating personalized prompts, which could significantly benefit a more diverse and more responsive audience, including those facing mental health challenges. By validating our method in a controlled environment, we are setting the stage to expand our study to wider populations, exploring the broader impact of journaling in technology-enhanced mental health interventions.

## 5 ETHICAL CONSIDERATIONS

Our study prioritizes high ethical standards to safeguard participant rights and well-being. Participants will give informed consent before starting the study, with the option to withdraw at any time. Data privacy is ensured through anonymization using individual IDs, secure storage, and restricted access to data. We advise participants to omit personal identifiers in journal entries and clarify that data is not monitored live, with emergency services information provided. The journaling screen displays a reminder about this. We will implement best practices for data security, allow participants to report prompt-related issues, and ensure GPT-4 generated prompts are free from sensitive content using a keyword filter. Participants will also be free to skip any journal entries as they choose. In addition, the data being passed on to GPT-4 is de-identified and contains high-level metadata only.

## 6 CONCLUSION

This study takes important strides in merging mobile technology with mental health practices by introducing a context-aware journaling application that extends the benefits of traditional reflection. Utilizing behavioral data collected via smartphones, and applying an LLM for prompt generation, we have crafted a system that provides personalized prompts on Android devices—a level of customization that goes beyond current applications. As this research unfolds, its findings may offer new directions for individualized mental health interventions and showcase the potential of technology-assisted self-help tools that can be integrated seamlessly into daily life, particularly for those in stressful academic environments and beyond.